\documentclass[preprintnumbers,amsmath,amssymb,superscriptaddress,pra,nofootinbib]{revtex4}
\usepackage{amsmath, amsthm, amsfonts, amssymb}
\usepackage{graphicx}
\usepackage{caption}
\usepackage{subcaption}
\usepackage{float}
\usepackage{xcolor}
\usepackage{mdframed}
\newtheorem{definition}{Definition}
\newtheorem{theorem}{Theorem}
\newtheorem{lemma}{Lemma}
\newtheorem{proposition}{Proposition}
\newtheorem{corollary}{Corollary}
\newtheorem{remarkenv}{Remark}

\newenvironment{remark}[1][]{%
    \smallskip
    \begin{mdframed}[
        topline=false,
        rightline=false,
        bottomline=false,
        leftline=true,
        linecolor=gray!60,
        linewidth=3pt,
        innerleftmargin=12pt,
        innerrightmargin=0pt,
        innertopmargin=4pt,
        innerbottommargin=4pt,
        backgroundcolor=gray!5
    ]%
    \if\relax\detokenize{#1}\relax
        \begin{remarkenv}%
    \else
        \begin{remarkenv}[#1]%
    \fi
}{%
    \end{remarkenv}%
    \end{mdframed}
    \smallskip
}

\newcommand{\xg}{\bar{x}_{\gamma}}
\newcommand{\yg}{\bar{y}_{\gamma}}
\newcommand{\zg}{\bar{z}_{\gamma}}
\newcommand{\pg}{\bar{p}_{\gamma}}

\begin{document}

\title{Generalizing Carleman linearization: nonlinear ODE systems as infinite linear ones}

\author{Andres M. Kowalski}
\email{kowalski@fisica.unlp.edu.ar}
\affiliation{Departamento de F\'{\i}sica-IFLP, Universidad Nacional de La Plata, La Plata, Argentina \\ and Comisi\'{o}n de Investigaciones Cient\'{i}ficas (CIC), Argentina}

\author{Ignacio S. Gomez}
\email{ignacio.gomez@uesb.edu.br}
\affiliation{Departamento de Ciências Exatas e Naturais, Universidade Estadual do Sudoeste da Bahia, BR 415, Itapetinga - BA, 45700-000, Brazil}
\affiliation{PROFÍSICA – Programa de Pós-Graduação em Física,  Universidade Estadual de Santa Cruz, 45650-000 Ilhéus, BA, Brazil}

\author{Gaspar Gonzalez Acosta}
\email{gaspar.gonzalez@fisica.unlp.edu.ar}
\affiliation{Departamento de F\'{\i}sica-IFLP, Universidad Nacional de La Plata, La Plata, Argentina \\ and Comisi\'{o}n de Investigaciones Cient\'{i}ficas (CIC), Argentina}

\begin{abstract}
We present a formalism that enables the solution of nonlinear ordinary differential equation (ODE) systems by reformulating them as extended systems of infinitely many linear equations. The additional variables are not restricted to monomials, as in standard Carleman linearization, but may be any family of functions closed under the Lie derivative.
This approach yields a local analytic solution expressed as a piecewise-defined function of time and initial conditions. For practical purposes, the method provides different types of approximations whose structure depends on the criteria adopted when introducing the new variables. Although all expansions share the Taylor coefficients of the solution, their partial sums may reveal distinct dynamical structures. The methodology is illustrated with paradigmatic examples: the Lorenz system, the quartic classical Hamiltonian, and the nonlinear pendulum. Finally, we briefly discuss the role of nonlinearity as a necessary condition (in finite dimension) for the emergence of chaos.
\end{abstract}

\maketitle

\section{Introduction}\label{sec:intro}
Differential equations have historically occupied a central role in science, appearing in diverse contexts such as mathematics, physics, biology, and economics. Typically, quantities of interest, their derivatives, and functions characterizing a system---depending on a single variable, time---are related through systems of ordinary differential equations (ODEs)~\cite{Coddington}. In mathematics and physics, the study of ODEs intersects significantly with dynamical systems and chaos theory~\cite{tabor, lich}. The mathematical structure of ODEs and the physical behavior they describe form one of the most established connections, both theoretically and numerically.

The distinction between linear and nonlinear ODEs is crucial: finite-dimensional linear systems exhibit regular behavior, while nonlinear systems can display complex and chaotic dynamics. Dissipative systems with strange attractors, such as the Lorenz attractor~\cite{Lorenz} and the R\"{o}ssler attractor~\cite{Rossler}, are among the most studied examples. Yet, the boundary between chaos and regularity is subtle. Mixed dynamical systems, for instance, may exhibit regular behavior in certain regions (the so-called islands of stability) and chaos in others.

Linear ODE systems with constant coefficients always admit exact well-known solutions, whereas nonlinear systems possess closed-form solutions only in exceptional cases. Classical results such as the Poincar\'{e}--Bendixson theorem, which ensures regular behavior in autonomous planar systems~\cite{Coddington, Poinc}, and Oseledets' theorem, which provides the theoretical foundation for Lyapunov exponents~\cite{Oseledets}, offer valuable insights. Other important results include Brudno's theorem, relating algorithmic complexity to Kolmogorov--Sinai entropy, and Pesin's theorem, relating Kolmogorov--Sinai entropy to positive Lyapunov exponents~\cite{QSDT}. In practice, however, nonlinear systems must be studied through approximations or numerical methods. Complementarily, characterizations of chaotic behavior can be described in terms of the Frobenius--Perron and Koopman transfer operators, which connect the approach to equilibrium in statistical mechanics and the levels of the ergodic hierarchy by means of invariant ergodic measures~\cite{LasotaMackey,Baladi}. This approach belongs to the spectral theory of transfer operators, which typically studies the decay of correlations between densities (or equivalently, observables) through time decay rates given in terms of the eigenvalues of the transfer operators~\cite{Baladi}.

Whether nonlinearity is a necessary condition for chaos depends on the dimension of the phase space and on the definition of chaos. For finite-dimensional systems a linear flow cannot be chaotic, so nonlinearity is necessary. In infinite dimension, however, there are linear operators that are chaotic in the sense of Devaney (hypercyclic, with a dense set of periodic points)~\cite{godefroy,grosse}, although their orbits are unbounded, unlike those of a strange attractor. Whether such operators exist depends on the space considered: there are Banach spaces that admit no chaotic operator~\cite{Bonet}.

In this work, we seek solutions of nonlinear ODE systems by reformulating them as infinite linear systems. The method involves introducing new variables corresponding to nonlinear functions in the original system and their successive derivatives. This is a natural framework, which goes back to Carleman~\cite{carleman1932,kowalskisteeb1991}, and is closely related to the Koopman operator approach~\cite{koopman1931,williams2015}. For Carleman linearization, explicit truncation error bounds~\cite{carlemanbounds}, Fourier-based extensions~\cite{carlemanfourier}, and globalization techniques for systems with several fixed points~\cite{carlemanglobal} have been developed, and it has been noted that the infinite linear system retains properties of the nonlinear one, including chaotic behavior~\cite{carlemanradio2016}. It has also been pointed out that chaotic systems do not admit exact finite-dimensional linear representations because of their mixed or continuous Koopman spectrum~\cite{koopmanchaosspectrum}. Carleman linearization has also been reviewed recently in Ref. \cite{MS25}, where linearization about equilibria and about the initial condition are compared and a pattern in the Taylor coefficients of the approximants is reported; a Carleman embedding of the quartic anharmonic oscillator, solved with Fourier series, goes back to Montroll and Helleman \cite{MH76}.

The expansion of the flow of an ODE system in powers of the initial conditions is also not new. Differential algebra techniques~\cite{berz} allow the automatic high-order expansion of the flow with respect to the initial conditions, and have been used to estimate the domain of validity of the expansion from the size of its coefficients~\cite{galileo} and to construct fast chaos indicators~\cite{lagrangiandescriptorsDA}. The validity of truncated Taylor expansions of chaotic maps has also been studied through the Lyapunov exponents of the truncated map as a function of the order~\cite{taylorchaoticmaps}.

The formalism presented here was developed independently of these approaches and differs from them in two respects, while placing a different emphasis on a third point. First, the new variables can be chosen according to different rules, depending on the problem (for instance Taylor in time, monomials of the original variables, or products of trigonometric functions, among many other possibilities), which yield different partial sums and reveal different dynamical structures. Second, the solution is built piecewise by re-expanding at successive times, yielding a local analytic representation on each subinterval (with local control by the Taylor remainder in the case of Rule~1). Third, and related to the expansions in the initial conditions mentioned above, we emphasize how the solution depends on the initial conditions---linearly for finite-dimensional linear systems, and via an infinite series in powers of the initial conditions for nonlinear ones---as a structural feature that is necessary for chaos in finite dimension, though not sufficient (for instance, integrable nonlinear systems such as the nonlinear pendulum also generate all powers of the initial conditions while remaining regular).

Section~\ref{sec:formalism} presents the formalism. Section~\ref{sec:applications} illustrates qualitative applications. Section~\ref{sec:chaos} discusses the role of nonlinearity as a condition for chaos, and conclusions are drawn in Section~\ref{sec:conclusions}. The Appendices provide properties of the exponential matrix and details on the numerical estimation of the Lyapunov exponent.

\section{Reformulation of a nonlinear ODE system as an infinite linear one}\label{sec:formalism}
We seek the analytic solution $\mathbf{x}(\mathbf{x}_0,t)$ of a nonlinear ODE system
\begin{equation}\label{sys equat}
\frac{d\,\mathbf{x}(t)}{dt} = \mathbf{f}(\mathbf{x}(t),t)
\end{equation}
(we denote this solution as $\mathbf{x}(t)$ for simplicity, assuming the dependence on the initial conditions), where $\mathbf{x}(t)$ and $\mathbf{f}(\mathbf{x}(t),t)$ are vector functions of $N$ components, rewriting it as an extended linear ODE system
\begin{equation}\label{sys lineal}
\frac{d\,\mathbf{X}(t)}{dt} = \mathbf{A}(t)\,\mathbf{X}(t).
\end{equation}
Here $\mathbf{X}(t)$ is a vector function of infinitely many components and $\mathbf{A}(t)$ is a square infinite-dimensional matrix with possibly time-dependent coefficients. The vector $\mathbf{X}(t)$,
\begin{equation}
\mathbf{X}(t) = \begin{pmatrix}
    \mathbf{x}(t) \\
    \mathbf{x}_E(t)
\end{pmatrix},
\label{X}
\end{equation}
includes $\mathbf{x}(t)$ as a subvector (as its first $N$ components, $X_j=x_j$ for $j\le N$). The remaining components $X_j$, $j>N$, form an infinite-dimensional vector $\mathbf{x}_E(t)$ of functions of $(\mathbf{x}(t),t)$. This construction is related to Carleman linearization~\cite{carleman1932,kowalskisteeb1991}, which also embeds a nonlinear system into an infinite linear one using monomials of the original variables; here the extended variables are chosen according to different rules (see Sec.~\ref{sec:applications}).

Let us take into account that the components of $\mathbf{f}(\mathbf{x}(t),t)$ can be written as a linear combination of the original variables and of new extended variables, that is,
\begin{equation}
\frac{dX_i}{dt}=f_i(\mathbf{x},t) = \sum_{j=1}^{m_1} a_{i,j}(t)\,X_j, \quad i = 1,\ldots,N,
\end{equation}
with $m_1>N$. Considering the functions $X_j$, $N<j\le m_1$, as new variables and repeating the procedure for the functions appearing in their time derivatives gives
\begin{equation}
\frac{d X_i}{dt} = \sum_{j=1}^{m_2} a_{i,j}(t)\,X_j, \quad i = N+1,\ldots,m_1,
\end{equation}
and so on, with $N<m_1<m_2<\cdots$. The coefficients $a_{i,j}(t)$ are the entries of the matrix $\mathbf{A}(t)$ in Eq.~\eqref{sys lineal}.

In this way, we construct the linear system of infinitely many equations~\eqref{sys lineal}. Different rules may be applied to select linear combinations, yielding distinct sets of variables $\mathbf{x}_E(t)$ and their associated matrices $\mathbf{A}(t)$. Clearly, the components of $\mathbf{x}_E(t)$ are not independent; rather, they constitute nonlinear functions of $(\mathbf{x}(t),t)$, a dependence encoded in their consistent initial values $\mathbf{X}(t_0)$.

For any finite-dimensional truncation $\mathbf{A}_{(n)}$ (or whenever $\mathbf{A}(t)$ is a continuous family of bounded linear operators on a Banach space\footnote{For a continuous family of bounded linear operators $\mathbf{A}(t)$ on a Banach space (in particular, for any finite-dimensional truncation $\mathbf{A}_{(n)}$), the Picard integral operator $(\mathcal{T}\mathbf{X})(t)=\mathbf{X}(t_0)+\int_{t_0}^{t}\mathbf{A}(s)\mathbf{X}(s)\,ds$ is a local contraction, guaranteeing a unique local fixed point and hence a unique local solution via the Contraction Mapping Principle~\cite{Rudin,Burton}. Moreover, the exponential of a bounded operator on a Banach algebra is defined by symbolic calculus~\cite{Rudin} and satisfies $\frac{d}{dt}\big(e^{\int_{t_0}^{t}\mathbf{A}\,ds}\mathbf{X}(t_0)\big)=\mathbf{A}\,e^{\int_{t_0}^{t}\mathbf{A}\,ds}\mathbf{X}(t_0)$ when $\mathbf{A}(t)$ commutes with $\int_{t_0}^{t}\mathbf{A}\,ds$. For the infinite system generated by nonlinear rules, $\mathbf{A}$ is generally unbounded, and the series representation is instead established directly via real analyticity in Theorem~\ref{thm:main} and Appendices~\ref{app:welldef} and~\ref{app:exist}.}), the truncated linear system $d\mathbf{X}_{(n)}/dt=\mathbf{A}_{(n)}(t)\mathbf{X}_{(n)}$ admits the unique solution
\begin{equation}\label{sol lineal}
\mathbf{X}_{(n)}(t) = \mathrm{T}\,e^{\int_{t_0}^t \mathbf{A}_{(n)}(s)\,ds}\,\mathbf{X}_{(n)}(t_0),
\end{equation}
where $\mathrm{T}$ is the time-ordering operator. Here $\mathbf{X}_{(n)}$ denotes the solution of the truncated system, which in general differs from the first components of the infinite vector $\mathbf{X}$, since couplings to the omitted variables are discarded.

We focus on the case of constant $\mathbf{A}$. This does not require $\mathbf{f}$ in Eq.~\eqref{sys equat} to be autonomous, provided that the explicit time dependence can be absorbed into $\mathbf{x}_E$ through auxiliary functions of $t$ that close under $L$ (Proposition~\ref{prop:rules}(b)). However, in all examples considered in this work, $\mathbf{f}$ is explicitly independent of time. Under these assumptions, Eq.~\eqref{sol lineal} reduces for finite truncations to
\begin{equation}\label{sol linea IMPORTANT}
\mathbf{X}_{(n)}(t) = e^{\mathbf{A}_{(n)}\,(t-t_0)}\,\mathbf{X}_{(n)}(t_0).
\end{equation}
In the general infinite-dimensional case, where $\mathbf{f}(\mathbf{x}(t),t)$ and subsequent derivatives are expressed as linear combinations $\sum_j a_{ij}\,g_j(\mathbf{x}(t),t)$ of functions $g_j$ of the state, the vector $\mathbf{X}(t)=\mathbf{g}(\mathbf{x}(t),t)$ satisfies~\eqref{sys lineal} componentwise. To extract the $N$-component state vector $\mathbf{x}(t)$ from $\mathbf{X}(t)$, we define the $N \times \infty$ reduced exponential matrix $\big[e^{\mathbf{A}(t-t_0)}\big]_R$ formed by the first $N$ rows of $e^{\mathbf{A}(t-t_0)}$. We now formalize this construction.

\begin{definition}[$L$-closed family]\label{def:closed}
Let $L=\partial_t+\mathbf{f}\cdot\nabla_{\mathbf{x}}$ be the Lie derivative along the flow of~\eqref{sys equat}, acting on functions of $(\mathbf{x},t)$. A family $\mathbf{g}=(g_1,g_2,\dots)$ of $C^\infty$ functions near $(\mathbf{x}_0,t_0)$ is \emph{$L$-closed} if
\begin{enumerate}
  \item[(a)] $g_i(\mathbf{x},t)=x_i$ for $i\le N$;
  \item[(b)] $L\,g_i=\sum_j a_{ij}\,g_j$ identically near $(\mathbf{x}_0,t_0)$, with constant $a_{ij}$ and finitely many nonzero entries for each $i$.
\end{enumerate}
We set $\mathbf{A}=(a_{ij})$ and $\mathbf{X}(t)=\mathbf{g}(\mathbf{x}(t),t)$, so that $\mathbf{X}(t_0)=\mathbf{g}(\mathbf{x}_0,t_0)$ is determined by $\mathbf{x}_0$ (\emph{consistent initial data}).
\end{definition}

\begin{theorem}\label{thm:main}
Let $\mathbf{f}(\mathbf{x},t)$ be real analytic jointly in $(\mathbf{x},t)$ near $(\mathbf{x}_0,t_0)$, let $\mathbf{x}(t)$ be the solution of~\eqref{sys equat} with $\mathbf{x}(t_0)=\mathbf{x}_0$, and let $\mathbf{g}$ be an $L$-closed family (Definition~\ref{def:closed}) with matrix $\mathbf{A}$ and $\mathbf{X}(t)=\mathbf{g}(\mathbf{x}(t),t)$. Then:
\begin{enumerate}
  \item[(i)] $\dfrac{d\mathbf{X}}{dt}=\mathbf{A}\mathbf{X}$ componentwise;
  \item[(ii)] (\emph{formal identity}) every power $\mathbf{A}^m$ is well defined and row-finite, and for all $m\ge0$
  \begin{equation}
       \mathbf{x}^{(m)}(t_0)=\big[\mathbf{A}^m\mathbf{X}(t_0)\big]_{1:N};
  \end{equation}
  \item[(iii)] (\emph{convergence}) there is $r(\mathbf{x}_0)>0$ such that, for $|t-t_0|<r(\mathbf{x}_0)$,
  \begin{equation}\label{Taylorfinal}
      \mathbf{x}(t)=\sum_{m=0}^{\infty}\frac{(t-t_0)^m}{m!}\big[\mathbf{A}^m\mathbf{X}(t_0)\big]_{1:N}
    = \big[e^{\mathbf{A}(t-t_0)}\big]_R\,\mathbf{X}(t_0),
  \end{equation}
  where $\big[e^{\mathbf{A}(t-t_0)}\big]_R\,\mathbf{X}(t_0)$ is defined by the series on the left-hand side (see Appendix~\ref{app:welldef} for its blockwise matrix representation).
\end{enumerate}
\end{theorem}

\begin{proof}
For any $C^\infty$ function $h(\mathbf{x},t)$, the chain rule gives $\frac{d}{dt}h(\mathbf{x}(t),t)=(Lh)(\mathbf{x}(t),t)$. Applied to $g_i$ this yields~(i).

(ii) Since $\mathbf{A}$ is row-finite, each entry of $\mathbf{A}^m$ is a finite sum, so $\mathbf{A}^m$ is well defined and row-finite. From $L\mathbf{g}=\mathbf{A}\mathbf{g}$ and the linearity of $L$ we get $L^m\mathbf{g}=\mathbf{A}^m\mathbf{g}$ by induction. Iterating the chain rule, $\mathbf{x}^{(m)}(t)=(L^m\mathbf{x})(\mathbf{x}(t),t)$, and since $g_i=x_i$ for $i\le N$, evaluating at $t_0$ gives $\mathbf{x}^{(m)}(t_0)=[\mathbf{A}^m\mathbf{g}]_{1:N}(\mathbf{x}_0,t_0)=[\mathbf{A}^m\mathbf{X}(t_0)]_{1:N}$. This step only uses that $\mathbf{f}$ and $\mathbf{g}$ are $C^\infty$.

(iii) Since $\mathbf{f}$ is real analytic, the solution of~\eqref{sys equat} is real analytic at $t_0$ (Cauchy's theorem for analytic ODEs), so its Taylor series converges to $\mathbf{x}(t)$ for $|t-t_0|<r(\mathbf{x}_0)$, for some $r(\mathbf{x}_0)>0$. By~(ii) its coefficients are those displayed, and by definition of the reduced exponential matrix this equals $\big[e^{\mathbf{A}(t-t_0)}\big]_R\,\mathbf{X}(t_0)$. Analyticity of $\mathbf{g}$ is not needed.
\end{proof}

A \emph{rule} is a prescription for choosing the family $\mathbf{g}$, i.e., the extended variables $\mathbf{x}_E$. The rules are not alternatives that must all work for every system: for each problem one may choose among those that are available, according to convenience. Theorem~\ref{thm:main} applies to any rule whose family is $L$-closed, and which rules are available depends on $\mathbf{f}$. Rule~1 is always available; Rules~2 and~3 are available when the corresponding monomials close under $L$ (Proposition~\ref{prop:rules}).

\begin{proposition}[Closure of the rules]\label{prop:rules}
Let $\mathbf{f}$ be real analytic jointly in $(\mathbf{x},t)$.
\begin{enumerate}
  \item[(a)] \emph{Rule 1}: $\mathbf{g}=(\mathbf{x},L\mathbf{x},L^2\mathbf{x},\dots)$ is $L$-closed for every such $\mathbf{f}$. $\mathbf{A}$ has identity blocks on the block superdiagonal and zeros elsewhere, and $\mathbf{X}=(\mathbf{x},\mathbf{x}',\mathbf{x}'',\dots)$.
  \item[(b)] \emph{Rule 2} (monomials): let $\mathbf{u}=(u_1,\dots,u_\nu)$ contain $\mathbf{x}$ and the auxiliary functions of $t$ needed to absorb explicit time dependence (e.g., $t$, $\cos\Omega t$, $\sin\Omega t$). If each $L\,u_k$ is a polynomial in $\mathbf{u}$ (in particular if $\mathbf{f}$ is polynomial in $\mathbf{x}$ with constant coefficients), the monomials of $\mathbf{u}$, including the constant $1$ (which may be omitted if no $L\,u_k$ has a constant term), form an $L$-closed family.
  \item[(c)] \emph{Rule 3} (products of other functions): let $\mathbf{u}=(u_1,\dots,u_\nu)$ be functions such that each $L\,u_k$ and each $L\,x_i=f_i$ is polynomial in $\mathbf{u}$. Then the monomials of $\mathbf{u}$, together with $\mathbf{x}$ itself, form an $L$-closed family. For example, for the nonlinear pendulum in Sec.~\ref{sec:pendulum} with $\mathbf{u}=(\sin x,\cos x,y)$, this family consists of the monomials of $\mathbf{u}$ together with $x$ (and the constant $1$, which can be omitted since no $L\,u_k$ contains a constant term). The monomials of $\mathbf{u}$ need not be linearly independent (e.g., $\sin^2x+\cos^2x=1$); this does not affect Theorem~\ref{thm:main}, but dimension counts below refer to the listed monomials.
\end{enumerate}
\end{proposition}

\begin{proof}
(a) $L(L^k\mathbf{x})=L^{k+1}\mathbf{x}$, so closure holds, and $L^k\mathbf{x}$ is analytic because $\mathbf{f}$ is.

(b),(c) Since $L$ is a derivation, for any multi-index $\boldsymbol{\alpha}=(\alpha_1,\dots,\alpha_\nu)$
\begin{equation}
  L(\mathbf{u}^{\boldsymbol{\alpha}})=\sum_{k=1}^\nu\alpha_k\,\mathbf{u}^{\boldsymbol{\alpha}-\mathbf{e}_k}\,L\,u_k,
\end{equation}
which is a finite combination of monomials whenever each $L\,u_k$ is polynomial in $\mathbf{u}$. If $\mathbf{u}=\mathbf{x}$ and $f_k=\sum_{\boldsymbol{\kappa}} c_{k,\boldsymbol{\kappa}}\mathbf{x}^{\boldsymbol{\kappa}}$ has degree $d$, then $L\mathbf{x}^{\boldsymbol{\alpha}}=\sum_{i,\boldsymbol{\kappa}}\alpha_ic_{i,\boldsymbol{\kappa}}\mathbf{x}^{\boldsymbol{\alpha}-\mathbf{e}_i+\boldsymbol{\kappa}}$ has degree at most $|\boldsymbol{\alpha}|+d-1$. If, moreover, $\mathbf{f}$ has no constant term (i.e., $c_{i,\boldsymbol{\kappa}}=0$ for $|\boldsymbol{\kappa}|=0$), the degree is also at least $|\boldsymbol{\alpha}|$, so $\mathbf{A}$ is block upper triangular with band width $d-1$. For the pendulum: $L\sin x=y\cos x$, $L\cos x=-y\sin x$, $L\,y=-\sin x$, $L\,x=y$, and $L\,1=0$.
\end{proof}

\begin{corollary}\label{cor:rules}
For every $L$-closed family, in particular for Rules~1--3 under the hypotheses of Proposition~\ref{prop:rules}, the series $[e^{\mathbf{A}(t-t_0)}]_R\mathbf{X}(t_0)$ reproduces the Taylor series of the solution of~\eqref{sys equat} and converges to it near $t_0$. Hence, whichever admissible rule is chosen for a given system, the resulting infinite Taylor series is the same; the rules differ in the structure of the matrix $\mathbf{A}$ and in their finite truncations (partial sums), which is what makes the choice a matter of convenience.
\end{corollary}

\begin{remark}\label{rem:closure}
Closure is not automatic. For the pendulum, monomials of $(x,y)$ do not close under $L$, because $L\,y=-\sin x$ is not polynomial; this is why a basis involving sines and cosines (Rule~3) is needed. A family that does not close must be completed with additional functions before the theorem can be applied.
\end{remark}

\begin{lemma}[Truncation and Taylor coefficients]\label{lem:coef}
Let $\mathbf{g}$ be an $L$-closed family graded by degree $\ell=1,2,\dots$ as in Appendix~\ref{app:struct}, so that $\mathbf{A}$ is block upper triangular (this requires the assumption stated there, i.e., that $L$ does not lower the degree; for monomial bases it holds when $\mathbf{f}$ has no constant term), and let $\mathbf{A}_{(n)}=\mathbf{A}^{[1,n]}$ be the truncation to degrees $\le n$, with $\mathbf{X}_{(n)}(t_0)$ the corresponding subvector of $\mathbf{X}(t_0)$. Let $M(n)$ be the largest $m$ such that, for every $m'\le m$, the first $N$ rows of $\mathbf{A}^{m'}$ involve only variables of degree $\le n$. Then, for $m\le M(n)$,
\begin{equation}
    \big[\mathbf{A}_{(n)}^m\mathbf{X}_{(n)}(t_0)\big]_{1:N}=\big[\mathbf{A}^m\mathbf{X}(t_0)\big]_{1:N}
  =\mathbf{x}^{(m)}(t_0).
\end{equation}
Hence the truncated approximant $\mathbf{x}_{(n)}(t)=\big[e^{\mathbf{A}_{(n)}(t-t_0)}\big]_R\mathbf{X}_{(n)}(t_0)$ has its first $M(n)+1$ Taylor coefficients exact, and each coefficient stabilizes at the Taylor value as $n\to\infty$. For monomial bases of a polynomial system of degree $d\ge2$, $M(n)=\lfloor(n-1)/(d-1)\rfloor$; in particular, $M(n)=K$ when $n=1+K(d-1)$. For $d=2$ this reproduces the pattern observed numerically in [MS] for the Taylor coefficients of Carleman approximants.
\end{lemma}

\begin{proof}
By the block upper triangular structure (Appendix~\ref{app:welldef}, Eq.~\eqref{eq:section}), the $(\ell,\ell')$ block of $\mathbf{A}^m$ is a sum over chains $\ell\le k_1\le\dots\le k_{m-1}\le \ell'$ involving only blocks $\mathbf{A}_{kk'}$ with $\ell\le k\le k'\le \ell'$. If the first $N$ rows of $\mathbf{A}^m$ involve only degrees $\le n$, these chains never leave $\mathbf{A}^{[1,n]}$, so $\mathbf{A}^m$ and $\mathbf{A}_{(n)}^m$ agree there. The last equality is Theorem~\ref{thm:main}(ii).
\end{proof}

By Theorem~\ref{thm:main}(iii), convergence holds on a finite interval $|t-t_0| < r(\mathbf{x}_0)$. In addition, one can construct local solutions on successive subintervals and concatenate them. As long as the solution exists and the local convergence radii $r_i$ remain bounded away from zero, a continuous succession of intervals can be formed, extending the solution in time. We thus express the solution of~\eqref{sys equat} as a piecewise-defined local analytic function
\begin{equation}\label{pegando}
\mathbf{x}(\mathbf{x}_0,t) = \big[e^{\mathbf{A}(t-t_{i-1})}\big]_R\,\mathbf{X}(t_{i-1}) \quad \text{if}\quad t_{i-1}\le t\le t_i, \quad i = 1,2,\ldots,
\end{equation}
with $\mathbf{x}(t_0)=\mathbf{x}_{0}$ and $\mathbf{X}(t_{i-1})=\mathbf{g}(\mathbf{x}(t_{i-1}),t_{i-1})$. In Eq.~\eqref{pegando}, the subintervals may have different lengths, and each step must satisfy $\Delta t_i = t_i - t_{i-1} < r_i(\mathbf{x}(t_{i-1}))$. In the numerical examples below the radii $r_i$ are not computed explicitly: the step is chosen following the heuristic criterion of Sec.~\ref{sec:lorenz}, and the condition $\Delta t_i<r_i$ is assumed rather than verified. Although the matrix $\mathbf{A}$ is the same in every interval, $r_i$ depends on the state vector $\mathbf{X}(t_{i-1})$ to which it is applied, so a formal proof of a uniform lower bound for the convergence radius remains open.

\begin{remark}[Scope of the results]\label{rem:scope}
Theorem~\ref{thm:main} and Lemma~\ref{lem:coef} concern the infinite object and the coefficients of the expansion. They do not assert that, for fixed $t$, $\mathbf{x}_{(n)}(t)\to\mathbf{x}(t)$ as $n\to\infty$: coefficient-wise convergence does not imply convergence of the partial sums without uniform control of the tails. For Rule~1 the partial sums are Taylor polynomials, so convergence within $|t-t_0|<r(\mathbf{x}_0)$ follows from Theorem~\ref{thm:main}(iii). For Rules~2 and~3 the validity of the approximation within each interval $\Delta t$, and in particular a uniform lower bound for the radius $r_i$, is assessed here only numerically, by comparison with a reference integration, and may be supported by the error bounds of Refs.~\cite{carlemanbounds,carlemanfourier,carlemanglobal}.
\end{remark}

Since $\mathbf{g}$ is $L$-closed and $\mathbf{f}$ is analytic, all derivatives of $\mathbf{x}(\mathbf{x}_0,t)$ at the beginning of each subinterval can be computed, yielding local approximate solutions whose error is controlled locally by the Taylor remainder (for $\Delta t_i < r_i$) in Rule~1; for Rules~2 and~3 see Remark~\ref{rem:scope}. This local control does not bound the accumulated error over long times in chaotic systems, where errors are amplified.

\emph{Initial conditions.} The solution $\mathbf{x}(\mathbf{x}_0,t)$ in Eqs.~\eqref{Taylorfinal} and~\eqref{pegando} also depends explicitly on powers of the initial conditions. Under the hypotheses of Theorem~\ref{thm:main}, the $i$-th component of $\mathbf{x}(\mathbf{x}_0,t)$ on the first interval (with analogous expressions on subsequent intervals) can be written as
\begin{equation}\label{fnolineal}
x_i(\mathbf{x}_0,t) = \sum_{m=0}^\infty \xi_{i,m}(\mathbf{x}_0)\,(t-t_0)^m, \quad i = 1,2,\ldots,N,
\end{equation}
where $\xi_{i,m}(\mathbf{x}_0)=\frac{1}{m!}\big[\mathbf{A}^m\mathbf{X}(t_0)\big]_i$ are nonlinear functions of the initial conditions $\mathbf{x}_0$, given as linear combinations of the components of $\mathbf{X}(t_0)$. Since $\mathbf{f}$ is real analytic, the solution is also real analytic in $\mathbf{x}_0$, so $x_i(\mathbf{x}_0,t)$ admits a joint Taylor series expansion in powers of $\mathbf{x}_0-\mathbf{x}_0^*$ (around a reference initial state $\mathbf{x}_0^*$; the choice $\mathbf{x}_0^*=\mathbf{0}$ gives the expansion about the origin) and $t-t_0$:
\begin{equation}\label{Tnolineal}
x_i(\mathbf{x}_0,t) = \sum_{|\boldsymbol{\mu}|=0}^\infty \sum_{m=0}^\infty C_{i,\boldsymbol{\mu},m}\,(\mathbf{x}_0-\mathbf{x}_0^*)^{\boldsymbol{\mu}}\,(t-t_0)^m, \quad i = 1,2,\ldots,N,
\end{equation}
where $\boldsymbol{\mu}=(\mu_1,\dots,\mu_N)\in\mathbb{N}_0^N$ is a multi-index with $|\boldsymbol{\mu}|=\sum_{r=1}^N \mu_r$ and $(\mathbf{x}_0-\mathbf{x}_0^*)^{\boldsymbol{\mu}} = (x_{0,1}-x^*_{0,1})^{\mu_1}\cdots(x_{0,N}-x^*_{0,N})^{\mu_N}$. This series converges in a neighborhood of $(\mathbf{x}_0^*,t_0)$, not necessarily on the whole domain of the solution. By contrast, if the system~\eqref{sys equat} is linear, homogeneous, and has constant coefficients, its solution reduces to
\begin{equation}\label{Tlineal}
x_i(\mathbf{x}_0,t) = \sum_{j=1}^N x_{0,j}\,\sum_{m=0}^\infty C_{i,j,m}\,(t-t_0)^m, \quad i = 1,2,\ldots,N,
\end{equation}
which represents the standard linear solution,
\begin{equation}\label{1}
x_i(\mathbf{x}_0,t) = \sum_{j=1}^N x_{0,j}\,\Phi_{ij}(t), \quad i = 1,2,\ldots,N,
\end{equation}
where $\Phi_{ij}(t)$ are linear combinations of exponentials of the eigenvalues of the linear system matrix, multiplied by polynomials in $t$.

From this point of view, the structural difference between the solution of a nonlinear ODE system and a finite-dimensional linear one is that the former entails a series expansion in higher-order powers of the initial conditions in addition to time, whereas the latter is strictly linear in the initial conditions (involving only first powers of $x_{0,j}$) multiplied by power series in time. Since a finite-dimensional linear flow cannot be chaotic, the appearance of higher-order powers of the initial conditions is a necessary ingredient for chaos in finite dimension; it is not by itself sufficient, as integrable nonlinear systems (such as the nonlinear pendulum in Sec.~\ref{sec:pendulum}) also generate all powers of $\mathbf{x}_0$ without exhibiting chaotic dynamics.

Although the exact solution is given locally by the piecewise temporal Taylor representation~\eqref{pegando}, practical computations rely on finite truncations determined by the choice of rule for $\mathbf{x}_E(t)$ in Eq.~\eqref{X}. One may truncate according to temporal derivative order (Rule~1), polynomial degree in state variables (Rule~2), or mixed trigonometric-polynomial bases (Rule~3, see Sec.~\ref{sec:pendulum}), among infinitely many theoretical possibilities. In all finite truncations, $\mathbf{A}_{(n)}$ and $\big[e^{\mathbf{A}_{(n)}(t-t_0)}\big]_R$ reduce to finite matrices of dimensions $S_n \times S_n$ and $N \times S_n$, respectively, where $n$ denotes the truncation order/degree and $S_n$ is the dimension of the truncated extended state vector. For Rule~1 we adopt the convention that the subindex $n$ counts blocks: Taylor order $K$ retains the $K+1$ blocks $\mathbf{x},\mathbf{x}',\dots,\mathbf{x}^{(K)}$, i.e., $n=K+1$ and $S_{K+1} = N(K+1)$. While all admissible infinite expansions reproduce the same Taylor series~\eqref{Taylorfinal}, their finite partial sums can display markedly different accuracy and dynamical features.

\section{Applications}\label{sec:applications}
In this section, we illustrate the main features of the methodology with three nonlinear examples: the chaotic Lorenz system, the quartic classical Hamiltonian, and the nonlinear pendulum. For each system, we consider the Lie derivative $L=\partial_t+\mathbf{f}\cdot\nabla_{\mathbf{x}}$ along the flow (Definition~\ref{def:closed}), construct an $L$-closed family $\mathbf{g}=(g_1,g_2,\dots)$ according to Proposition~\ref{prop:rules}, and obtain approximate trajectories of the $N$-dimensional state vector $\mathbf{x}(t)$ from finite truncations via Theorem~\ref{thm:main}(iii) and Lemma~\ref{lem:coef}. We write $\mathbf{g}_{(n)}$ for the truncated family (degrees $\le n$; $K+1$ blocks for Rule~1 at order $K$), $\mathbf{X}_{(n)}(t)=\mathbf{g}_{(n)}(\mathbf{x}(t))$ for the truncated extended vector, and $\mathbf{A}_{(n)}$ for the corresponding matrix. Because all three systems are autonomous and the selected basis functions $g_i(\mathbf{x})$ depend only on the state variables (with no explicit $t$-dependence), the partial derivative $\partial_t$ acts trivially ($\partial_t g_i = 0$) and $L$ reduces effectively to $\mathbf{f}\cdot\nabla_{\mathbf{x}}$. The analysis is primarily qualitative and intended for illustration, complemented by selected error and dynamical benchmarks; a comprehensive quantitative study will be presented in future work.

In all examples, the time interval $[0,t_{\max}]$ with $t_{\max}=10$ is covered by $N_{\mathrm{steps}}=t_{\max}/\Delta t$ subintervals of equal length $\Delta t=0.01$. The $\gamma$-th subinterval ($\gamma=1,\dots,N_{\mathrm{steps}}$) covers the global time range $[(\gamma-1)\Delta t,\gamma\Delta t]$, and $\tau\in[0,\Delta t]$ denotes the local time, so that the global time is $t=(\gamma-1)\Delta t+\tau$. The state at the beginning of the $\gamma$-th subinterval is denoted $\bar{\mathbf{x}}_\gamma=\mathbf{x}((\gamma-1)\Delta t)$, with $\bar{\mathbf{x}}_1=\mathbf{x}_0$ (components $\bar{x}_\gamma,\bar{y}_\gamma,\bar{z}_\gamma$ or $\bar{p}_\gamma$ in each example), and the corresponding extended vector is $\bar{\mathbf{X}}_{(n),\gamma}=\mathbf{g}_{(n)}(\bar{\mathbf{x}}_\gamma)$. Each subinterval uses as its initial state the approximate state obtained at the end of the previous one (see Eq.~\eqref{pegando}). Grid times are denoted $t_k=k\Delta t$.

\subsection{Lorenz system}\label{sec:lorenz}

The Lorenz system represents chaotic dynamics par excellence in the literature on dynamical systems. For certain parameter values and initial conditions, its solutions give rise to the so-called Lorenz attractor, which resembles a butterfly when plotted. Due to its strong sensitivity to initial conditions, this system is ideal for testing chaotic behavior. It is defined by the equations
\begin{equation}\label{lorenz}
\begin{aligned}
\dot{x} &= \sigma (y-x), \\
\dot{y} &= x(\rho-z)-y, \\
\dot{z} &= xy-\beta z,
\end{aligned}
\end{equation}
so that $\mathbf{x}=(x,y,z)^\top$ ($N=3$) and the associated Lie derivative (Definition~\ref{def:closed}) is
\begin{equation}
L = \partial_t + \sigma(y-x)\,\frac{\partial}{\partial x} + \big[x(\rho-z)-y\big]\,\frac{\partial}{\partial y} + (xy-\beta z)\,\frac{\partial}{\partial z},
\end{equation}
where $\partial_t$ vanishes when acting on functions of $\mathbf{x}$ alone. Nonlinearity arises from the second and third equations, while the first describes a linear relaxation of $x$ toward $y$ proportional to $y-x$. Within our framework, this feature is reflected in the fact that the initial conditions $\mathbf{X}(t_0)=\mathbf{g}(\mathbf{x}_0)$ of the extended variables are linked nonlinearly to $\mathbf{x}_0$.

\emph{Rule 1.} Following Proposition~\ref{prop:rules}(a), we first choose the $L$-closed family generated by repeated action of $L$ on the state variables, $\mathbf{g}=(\mathbf{x}, L\mathbf{x}, L^2\mathbf{x}, \dots)$. Since $\mathbf{x}$ and $\mathbf{f}(\mathbf{x})$ have no explicit time dependence, $\partial_t(L^{j}\mathbf{x})=0$ at every order, so $\mathbf{X}^{(j)}(t) = (L^j\mathbf{x})(\mathbf{x}(t)) = (\mathbf{f}\cdot\nabla_{\mathbf{x}})^j\mathbf{x}(t)$ defines the successive $N$-component derivative blocks
\begin{equation}
\mathbf{X}^{(0)}(t)=\mathbf{x}(t), \quad \mathbf{X}^{(1)}(t)=\mathbf{x}'(t), \quad \ldots, \quad \mathbf{X}^{(j)}(t)=\mathbf{x}^{(j)}(t), \quad j=0,1,2,\ldots
\end{equation}
Since $L(L^{j}\mathbf{x})=L^{j+1}\mathbf{x}$, this choice yields an infinite matrix $\mathbf{A}$ constructed with $N \times N$ blocks $\mathbf{A}_{kk'}=\delta_{k+1,k'}\mathbf{I}$, where $\mathbf{I}$ is the $N \times N$ identity matrix and $\delta_{k+1,k'}$ is the Kronecker delta. Truncating at Taylor order $K$ retains the $K+1$ blocks $(\mathbf{X}^{(0)},\dots,\mathbf{X}^{(K)})$, so that $\mathbf{A}_{(K+1)}$ has dimension $N(K+1)$ and $\mathbf{A}_{(K+1)}^{K+1}=\mathbf{0}$ (Appendix~\ref{app:nilp}). Projecting onto the first $N=3$ rows via $[e^{\mathbf{A}_{(K+1)} \tau}]_R\,\bar{\mathbf{X}}_{(K+1),\gamma}$ reproduces the $K$-th degree Taylor polynomial of $\mathbf{x}(t)$ on the $\gamma$-th interval.

\begin{figure}[H]
     \centering
     \begin{subfigure}{0.49\textwidth}
         \centering
         \includegraphics[width=\textwidth]{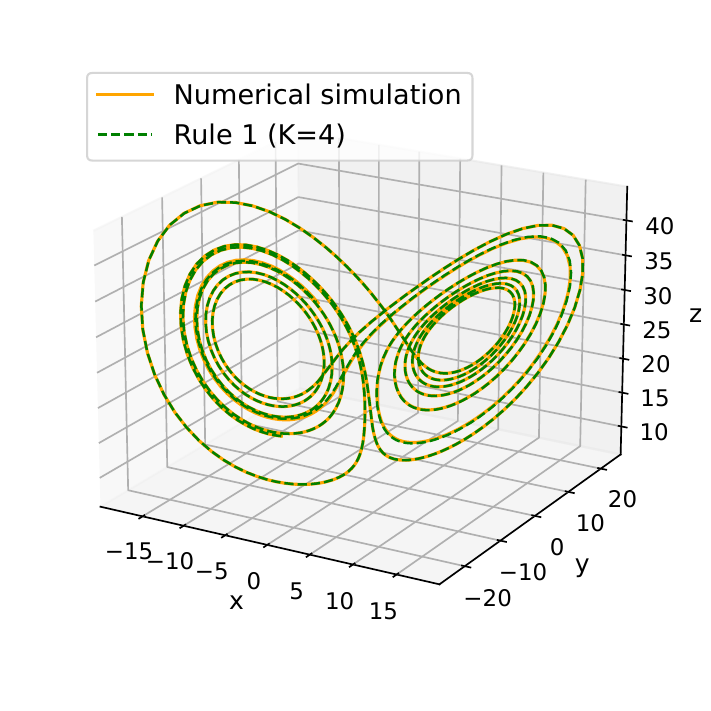}
         \caption{}
         \label{Lorenz_r1ord4}
     \end{subfigure}
     \begin{subfigure}{0.48\textwidth}
         \centering
         \includegraphics[width=\textwidth]{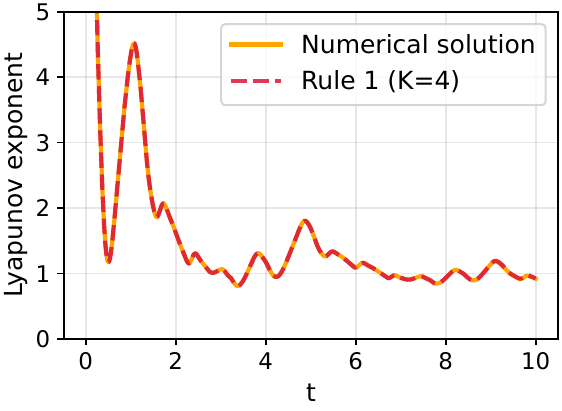}
         \caption{}
         \label{lyapunov_r1_ord4}
     \end{subfigure}
\caption{Lorenz model with parameters $\sigma=10$, $\rho=28$, $\beta=8/3$ and initial conditions $x(0)=-6$, $y(0)=-10$, $z(0)=14$, within the Lorenz attractor region over $[0,10]$: (a) parametric plot of $(x(t),y(t),z(t))$ for the numerical solution and Rule~1 ($K=4$), with a discrete $\ell^2$ relative error of $3.85\times10^{-2}$, and (b) finite-time largest Lyapunov characteristic exponent using the same parameters and initial conditions as in (a) together with the perturbed trajectory $x(0)=-6+10^{-8}$, $y(0)=-10+10^{-8}$, $z(0)=14+10^{-8}$.}
\label{Fig1}
\end{figure}

For the parameter values $\sigma=10$, $\rho=28$, $\beta=8/3$ and initial conditions $x(0)=-6$, $y(0)=-10$, $z(0)=14$ within the Lorenz attractor region, we compute the trajectory over the interval $[0,10]$. To achieve this, we concatenate consecutive local solutions of equal step length $\Delta t=0.01$ (the criterion used to choose this step is discussed below) and expand in powers of the local time $\tau \in [0,\Delta t]$ up to order $K=4$. Measured by the discrete $\ell^2$ relative error $\big(\sum_{k} \|\mathbf{x}_{\mathrm{approx}}(t_k)-\mathbf{x}_{\mathrm{num}}(t_k)\|_2^2 \big/ \sum_{k} \|\mathbf{x}_{\mathrm{num}}(t_k)\|_2^2\big)^{1/2}$ over the grid $t_k\in[0,10]$, the discrepancy between the Rule~1 ($K=4$) and reference numerical solutions is $3.85\times10^{-2}$. The results are shown in Fig.~\ref{Lorenz_r1ord4}, together with the numerical solution, where the characteristic ``butterfly'' of the Lorenz attractor is clearly observed. In Fig.~\ref{lyapunov_r1_ord4}, we plot the finite-time largest Lyapunov characteristic exponent for both the approximate and numerical trajectories over $[0,10]$, showing close agreement throughout the integration window. Further details on the calculation method are provided in Appendix~\ref{Apendice_Lya}.

To observe the structure of the expansion with Rule~$1$, we take the particular case of order $K=2$. For the $\gamma$-th interval, the local temporal evolution $\tau \in [0, \Delta t]$ is given by
\begin{equation}
\mathbf{x}((\gamma-1)\Delta t + \tau) = \bar{\mathbf{x}}_\gamma + \mathbf{f}(\bar{\mathbf{x}}_\gamma)\,\tau + \dfrac{1}{2}(L\mathbf{f})(\bar{\mathbf{x}}_\gamma)\,\tau^2.
\end{equation}
The explicit expansion for the $x$ component is
\begin{equation}
x((\gamma-1)\Delta t + \tau) = \xg + \sigma(\yg - \xg)\,\tau + \dfrac{1}{2}\sigma \left[ \xg(\rho - \zg) - \yg - \sigma(\yg - \xg) \right] \tau^2.
\end{equation}
Evaluating the local approximation at $\tau = \Delta t$ advances the state to the global time $\gamma \Delta t$. The same indexing procedure is applied in the examples of the subsequent rules.

\emph{Rule 2.} Next, following Proposition~\ref{prop:rules}(b) with $\mathbf{u}=\mathbf{x}=(x,y,z)^\top$ (and omitting the constant $1$ since $\mathbf{f}$ has no constant term), we construct the $L$-closed family $\mathbf{g}$ formed by all monomials $x^a y^b z^{\ell-a-b}$ ordered by total degree $\ell=1,2,\ldots$, with $(g_1,g_2,g_3)=(x,y,z)$. Because $\mathbf{f}$ is quadratic ($d=2$) and $\partial_t g_i = 0$, $L$ maps a monomial of degree $\ell$ into a linear combination of monomials of degrees $\ell$ and $\ell+1$, e.g., $L(x^2)=-2\sigma x^2+2\sigma xy$ and $L(xy)=\rho x^2-(1+\sigma)xy+\sigma y^2-x^2 z$, giving $\mathbf{A}$ a block upper-triangular structure of band width $d-1=1$. Truncating $\mathbf{g}$ at degree $n=2$ ($1 \leq \ell \leq 2$) and concatenating consecutive intervals of length $\Delta t=0.01$, the truncated $9 \times 9$ matrix $\mathbf{A}_{(2)}$ is
\begin{equation}\label{eq:A2_lorenz}
\mathbf{A}_{(2)}=\left(
  \begin{array}{ccccccccc}
    -\sigma & \sigma & 0 & 0 & 0 & 0 & 0 & 0 & 0 \\
    \rho & -1 & 0 & 0 & 0 & -1 & 0 & 0 & 0 \\
    0 & 0 & -\beta & 0 & 1 & 0 & 0 & 0 & 0 \\
    0 & 0 & 0 & -2\sigma & 2\sigma & 0 & 0 & 0 & 0 \\
    0 & 0 & 0 & \rho & -1-\sigma & 0 & \sigma & 0 & 0 \\
    0 & 0 & 0 & 0 & 0 & -\sigma-\beta & 0 & \sigma & 0 \\
    0 & 0 & 0 & 0 & 2\rho & 0 & -2 & 0 & 0 \\
    0 & 0 & 0 & 0 & 0 & \rho & 0 & -1-\beta & 0 \\
    0 & 0 & 0 & 0 & 0 & 0 & 0 & 0 & -2\beta
  \end{array}
\right),
\end{equation}
acting on the truncated vector $\mathbf{X}_{(2)}(t)=\mathbf{g}_{(2)}(\mathbf{x}(t))$,
\begin{equation}
\mathbf{X}_{(2)}(t)=\left(x, y, z, x^2, xy, xz, y^2, yz, z^2 \right)^\top,
\end{equation}
from which the $3$-dimensional trajectory $\mathbf{x}(t)=(x(t),y(t),z(t))^\top$ on the $\gamma$-th interval is recovered via $\mathbf{x}((\gamma-1)\Delta t+\tau)=[e^{\mathbf{A}_{(2)} \tau}]_R\,\bar{\mathbf{X}}_{(2),\gamma}$, with $\bar{\mathbf{X}}_{(2),\gamma}=\mathbf{g}_{(2)}(\bar{\mathbf{x}}_\gamma)$.

\begin{figure}[H]
     \centering
     \begin{subfigure}{0.49\textwidth}
         \centering
         \includegraphics[width=\textwidth]{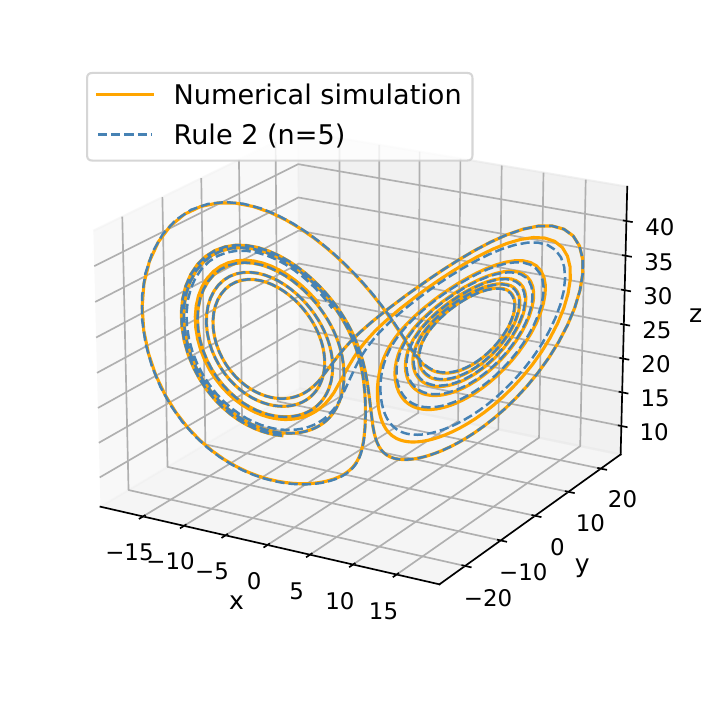}
         \caption{}
         \label{Lorenz_r2ord5}
     \end{subfigure}
     \begin{subfigure}{0.48\textwidth}
         \centering
         \includegraphics[width=\textwidth]{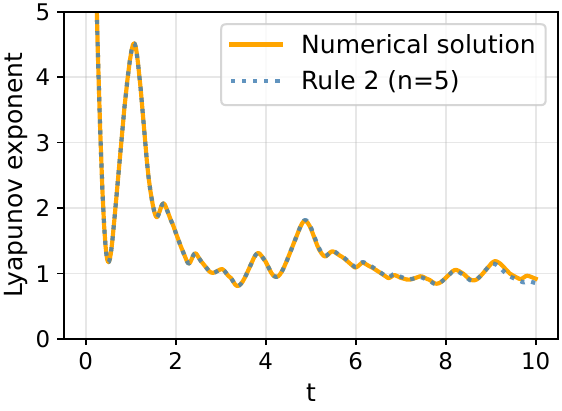}
         \caption{}
         \label{lyapunov_r2_ord5}
     \end{subfigure}
\caption{Parametric plot of $(x(t),y(t),z(t))$ and finite-time largest Lyapunov exponent for the Lorenz model over $[0,10]$ with the same parameters and initial conditions as in Fig.~\ref{Fig1}: (a) numerical solution and Rule~2 (degree $n=5$), with a discrete $\ell^2$ relative error (same definition as in Fig.~\ref{Fig1}) of $3.738\times10^{-1}$, and (b) finite-time largest Lyapunov characteristic exponent for the approximate and numerical trajectories.}
\label{Fig2}
\end{figure}

Figure~\ref{Fig2} shows the result of Rule~2 at degree $n=5$ over $[0,10]$ for the same parameters and initial conditions. Its discrete $\ell^2$ relative error is $3.738\times10^{-1}$, larger than the $3.85\times10^{-2}$ obtained with Rule~1 ($K=4$) in Fig.~\ref{Fig1}. The monomial basis of Rule~2 is centred at the equilibrium $\mathbf{x}=0$ (note that $\mathbf f(0)=0$), so Rule~2 coincides with Carleman linearization about an equilibrium (``Type~A'' in Ref. \cite{MS25}), whose accuracy is reported to degrade away from that equilibrium; the initial state used here is far from the origin, which may account for the difference. Linearizing instead about the current state at each step (``Type~B'' in Ref. \cite{MS25}) is reported to improve the accuracy; we do not explore this variant here. The comparison refers only to this trajectory and this step size.

We now present the analytic approximation for $x$ (the first component of $[e^{\mathbf{A}_{(2)} \tau}]_R\,\bar{\mathbf{X}}_{(2),\gamma}$) at degree $n=2$, for the chaotic parameter values $\sigma=10$, $\rho=28$, $\beta=8/3$, and general initial conditions. For the $\gamma$-th interval ($\tau \in [0,\Delta t]$), we obtain
\begin{equation}
\label{exponential}
\begin{aligned}
x((\gamma-1)\Delta t+\tau)=\;&e^{11.83\,\tau}\big[0.3702\,\xg +0.2886\,\yg-(0.0449\,\xg +0.0290\,\yg)\,\zg\big]\\
+\;&e^{-22.83\,\tau}\big[0.6299\,\xg-0.2886\,\yg+(0.0648\,\xg-0.0338\,\yg)\,\zg\big]\\
+\;&e^{9.16\,\tau}\,(0.0434\,\xg+0.0338\,\yg)\,\zg\\
+\;&e^{-25.49\,\tau}\,(-0.0633\,\xg+0.0290\,\yg)\,\zg.
\end{aligned}
\end{equation}
In this case, real exponential modes associated with the eigenvalues of $\mathbf{A}_{(2)}$ appear instead of powers of $\tau$ as in Rule~1, directly exposing the local unstable ($11.83$, $9.16$) and stable ($-22.83$, $-25.49$) exponential rates of the truncated flow.

To compare both rules under matched Taylor-order information content (Lemma~\ref{lem:coef}), the temporal order $K$ is related to the maximum degree $n$ of the monomial basis by
\begin{equation}
n = 1 + K(d - 1),\label{relat_regla12}
\end{equation}
where $d\ge2$ is the maximum polynomial degree of the differential system. For the Lorenz system, which is quadratic ($d=2$) due to the bilinear terms $xz$ and $xy$, this reduces to $n = K + 1$. With this choice, the coefficients of $\tau^m$ in Rule~2 coincide with those of the Taylor expansion for $m \le K$, while differences may arise from $\tau^{K+1}$ onward. Thus, at fixed $\Delta t$, one isolates the effect of truncating monomial degree versus temporal order. This relation provides a sufficient condition (worst-case bound) on Taylor coefficient matching rather than an equality of global errors.

The step size is guided by the spectrum of the truncated matrix $\mathbf{A}_{(n)}$ of Rule~2, with spectral abscissa $\lambda_{\max}=\max\mathrm{Re}\,\mathrm{eig}(\mathbf{A}_{(n)})=n\lambda_+$, where $\mathrm{eig}(\mathbf{A}_{(n)})$ denotes the set of eigenvalues of $\mathbf{A}_{(n)}$. For the Lorenz system, $\lambda_+\approx 11.83$, yielding $\lambda_{\max}\approx 23.66$ for $n=2$ (and $\lambda_{\max}\approx 59.15$ for $n=5$). Its inverse, $\tau_{\mathrm{c}}=1/\lambda_{\max}$, sets the timescale over which $\exp(\mathbf{A}_{(n)}\,\Delta t)$ amplifies the dominant mode by $e^{\lambda_{\max}\Delta t}$ per interval. We adopt $\Delta t<\tau_{\mathrm{c}}$ as a conservative heuristic criterion (satisfied by $\Delta t=0.01$), and use the same fixed step $\Delta t$ (without adaptive substeps) in Rule~1, so that both integrations operate on the same temporal mesh for direct comparison. This criterion concerns the amplification of the truncated Rule~2 flow per step (for Rule~1 the matrix $\mathbf{A}_{(K+1)}$ is nilpotent and has no such timescale), and it is not a bound on the convergence radii $r_i$.

\begin{figure}[H]\centering
\scalebox{.8}{\includegraphics{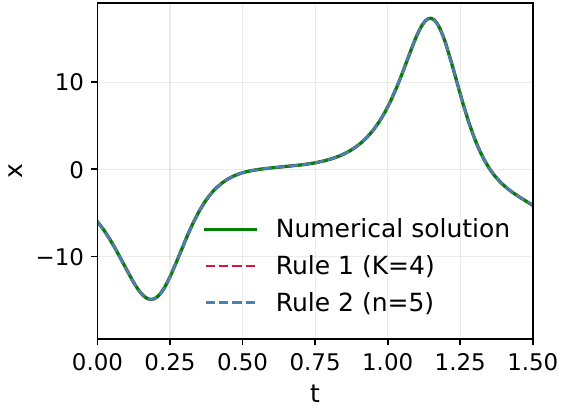}}
\caption{Lorenz model: short-time evolution of $x(t)$ obtained using Rules~1 and~2 (with $K=4$ and degree $n=5$, respectively), along with the reference numerical solution. Parameter values and initial conditions are the same as in Fig.~\ref{Fig1}.}
\label{Fig3}
\end{figure}

\begin{figure}[H]\centering
\scalebox{.8}{\includegraphics{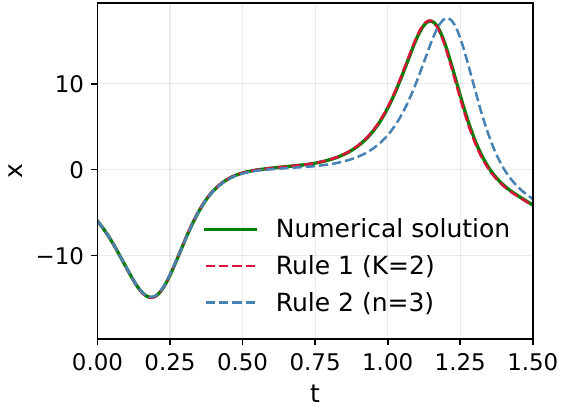}}
\caption{Lorenz model: short-time evolution of $x(t)$ obtained using Rules~1 and~2 (with $K=2$ and degree $n=3$, respectively), together with the numerical solution. Parameter values and initial conditions are as in Fig.~\ref{Fig1}.}
\label{Fig4}
\end{figure}

Figures~\ref{Fig3} and~\ref{Fig4} show the short-time evolution of $x(t)$ obtained with both approaches and compared against the reference numerical solution for $(K=4, n=5)$ and $(K=2, n=3)$, respectively. In both cases, the approximations accurately capture the short-time trajectory of $x(t)$.

\subsection{Quartic Classical Hamiltonian}\label{sec:quartic}
For a particle of mass $m$ in one dimension, the quartic classical Hamiltonian is expressed as
\begin{equation}
H = \dfrac{p^2}{2m} + \dfrac{1}{2}m\omega^2 x^2 + \eta x^4,
\label{eq:H_quartic}
\end{equation}
where $x$ and $p$ represent the position and linear momentum in phase space, respectively, and $\omega$ is the characteristic angular frequency of the unperturbed harmonic potential. The parameter $\eta \geq 0$ regulates the strength of the quartic anharmonicity of the potential. From Hamilton's canonical equations, the equations of motion for the dynamical state $\mathbf{x}(t)=(x(t),p(t))^\top$ ($N=2$) are given by
\begin{subequations}
\label{eq:hamilton_eqs}
\begin{align}
\dot{x} &= \frac{\partial H}{\partial p} = \frac{p}{m}, \\
\dot{p} &= -\frac{\partial H}{\partial x} = -m\omega^2 x - 4\eta x^3,
\end{align}
\end{subequations}
with the corresponding Lie derivative (Definition~\ref{def:closed})
\begin{equation}
L = \partial_t + \frac{p}{m}\,\frac{\partial}{\partial x} - \left(m\omega^2 x + 4\eta x^3\right)\frac{\partial}{\partial p},
\end{equation}
where $\partial_t$ again drops out when acting on time-independent functions of $(x,p)$. To avoid confusion with the mass $m$, the letter $\ell$ is used below for the degree of monomials and $j$ for derivative orders.

The system of differential equations~\eqref{eq:hamilton_eqs} is a polynomial system of maximum degree $d=3$ in $x$ and $p$. Therefore, the relation~\eqref{relat_regla12} between Rule~1 and Rule~2 reduces to $n = 1 + 2K$. Applying Proposition~\ref{prop:rules}(b), the $L$-closed family $\mathbf{g}$ for Rule~2 consists of all monomials $x^a p^{\ell-a}$ ($0 \le a \le \ell$) ordered by total degree $\ell=1,2,\ldots$, with $(g_1,g_2)=(x,p)$. Since $d=3$, $\mathbf{f}$ has no constant term, and $\partial_t g_k = 0$, $L$ maps monomials of degree $\ell$ into linear combinations of degrees $\ell$ and $\ell+2$, yielding a block upper-triangular matrix $\mathbf{A}$ of band width $d-1=2$.

Considering degree $n=3$ as the minimum truncation required to capture the cubic force in~\eqref{eq:hamilton_eqs}, we obtain the truncated vector $\mathbf{X}_{(3)}(t)=\mathbf{g}_{(3)}(\mathbf{x}(t))$,
\begin{equation}
    \mathbf{X}_{(3)}(t)=\left(x, p, x^2, xp, p^2, x^3, x^2p, xp^2, p^3 \right)^\top,
\end{equation}
whose associated $9 \times 9$ matrix $\mathbf{A}_{(3)}$ is
\begin{equation}
\mathbf{A}_{(3)} = \begin{pmatrix} 0 & \frac{1}{m} & 0 & 0 & 0 & 0 & 0 & 0 & 0 \\ -m\omega^2 & 0 & 0 & 0 & 0 & -4\eta & 0 & 0 & 0 \\ 0 & 0 & 0 & \frac{2}{m} & 0 & 0 & 0 & 0 & 0 \\ 0 & 0 & -m\omega^2 & 0 & \frac{1}{m} & 0 & 0 & 0 & 0 \\ 0 & 0 & 0 & -2m\omega^2 & 0 & 0 & 0 & 0 & 0 \\ 0 & 0 & 0 & 0 & 0 & 0 & \frac{3}{m} & 0 & 0 \\ 0 & 0 & 0 & 0 & 0 & -m\omega^2 & 0 & \frac{2}{m} & 0 \\ 0 & 0 & 0 & 0 & 0 & 0 & -2m\omega^2 & 0 & \frac{1}{m} \\ 0 & 0 & 0 & 0 & 0 & 0 & 0 & -3m\omega^2 & 0 \end{pmatrix}.
\end{equation}
Assuming unit values $m=1$, $\omega=1$, and $\eta=1$ for the model parameters and projecting $[e^{\mathbf{A}_{(3)} \tau}]_R\,\bar{\mathbf{X}}_{(3),\gamma}$ onto its first component for local time $\tau \in [0,\Delta t]$, the explicit analytic approximation for the position $x$ at truncation degree $n=3$ (the numerical results below use $n=7$) is obtained as
\begin{equation}
\begin{aligned} x((\gamma-1)\Delta t + \tau) =\;& \left[ \frac{8\xg - \xg^3 + 3\xg\, \pg^2}{16} + i \frac{9\pg^3 + 21\xg^2\, \pg - 8\pg}{16} + \frac{3\left(\xg^2 +\pg^2\right)}{4} (\pg + i \xg) \tau \right] e^{i\tau} \\ 
+\;& \left[ \frac{8\xg - \xg^3 + 3\xg\, \pg^2}{16} - i \frac{9\pg^3 + 21\xg^2\, \pg - 8\pg}{16} + \frac{3\left(\xg^2 + \pg^2\right)}{4} (\pg - i \xg)\tau \right] e^{-i\tau} \\ 
+\;& \left[ \frac{\xg^3 - 3\xg\, \pg^2}{16} + i \frac{\pg^3 - 3\xg^2\, \pg}{16} \right] e^{3i\tau} \\ 
+\;& \left[ \frac{\xg^3 - 3\xg\, \pg^2}{16} - i \frac{\pg^3 - 3\xg^2\, \pg}{16} \right] e^{-3i\tau}. \end{aligned}
\end{equation}
Since $x(t)$ is a real physical quantity, the complex-conjugate pairs combine via Euler's formula into a purely real trigonometric expression, explicitly revealing the oscillation frequencies of the system:
\begin{equation}
\begin{aligned}
x((\gamma-1)\Delta t + \tau)=\;&\frac{3}{2}\,\Big(\xg^2+\pg^2\Big)\,\tau\,\big[\pg\cos\tau-\xg\sin\tau\big]\\
&+\frac{8\xg+3\xg\, \pg^2-\xg^3}{8}\cos\tau+\frac{8\pg-9\pg^3-21\xg^2\, \pg}{8}\sin\tau\\
&+\frac{\xg^3-3\xg\, \pg^2}{8}\cos 3\tau+\frac{3\xg^2\, \pg-\pg^3}{8}\sin 3\tau.
\end{aligned}
\end{equation}

\begin{figure}[H]
\centering
  \scalebox{.8}{\includegraphics{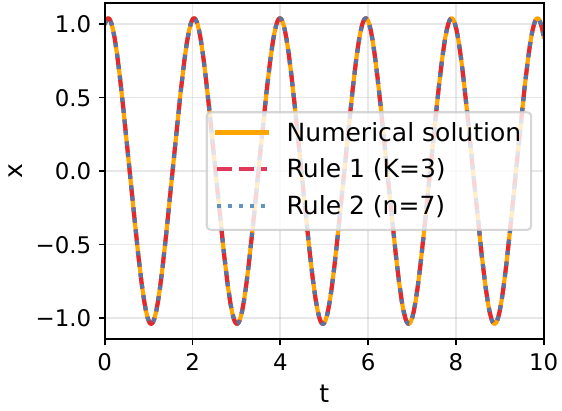}}
    \caption{Time evolution of $x(t)$ for the quartic classical Hamiltonian~\eqref{eq:H_quartic} over $[0,10]$ ($\Delta t = 0.01$) using Rule~1 ($K=3$) and Rule~2 ($n=7$), compared with the numerical solution. The initial conditions are $x(0)=1$ and $p(0)=0.5$.}
    \label{fig:hamilt_quart}
\end{figure}

Figure~\ref{fig:hamilt_quart} shows the time evolution of the position $x(t)$ over $t \in [0,10]$ with step size $\Delta t = 0.01$ obtained for the quartic classical Hamiltonian using $K=3$ in Rule~1 and the matched degree $n=7$ in Rule~2. Visually, the trajectories corresponding to the reference numerical solution, Rule~1, and Rule~2 appear indistinguishable. However, quantitative comparison shows that the maximum error relative to the peak amplitude, $\max_k |x_{\mathrm{approx}}(t_k)-x_{\mathrm{num}}(t_k)| / \max_k |x_{\mathrm{num}}(t_k)|$, associated with Rule~1 is $1.11 \times 10^{-2}$, whereas for Rule~2 it decreases to $9.46 \times 10^{-9}$.

Since the system is conservative (Hamiltonian), the total energy $H(x(t),p(t))$ must remain constant along the trajectory. When evaluating the maximum absolute energy fluctuation $\max_k |H(x(t_k),p(t_k)) - H(x(0),p(0))|$ over the integration interval $[0,10]$, Rule~1 exhibits a maximum variation of $1.68 \times 10^{-3}$, in contrast to $5.18 \times 10^{-10}$ for Rule~2. These results, obtained for a single initial condition and a single step size, suggest that Rule~2 can achieve substantially higher accuracy and better conservation of the energy for this system; note also that Rule~2 with $n=7$ uses a considerably larger basis than Rule~1 with $K=3$.

\subsection{Nonlinear pendulum}\label{sec:pendulum}
The nonlinear pendulum is one of the most widely studied systems in classical mechanics and dynamical systems theory, both physically and mathematically. Its importance lies in the universality of the physical behavior it embodies, while its simplicity allows for a complete characterization of its dynamics. Renaming its angle $\theta$ and angular velocity $\dot{\theta}$ as $x$ and $y$, respectively, the equations of motion for $\mathbf{x}(t)=(x(t),y(t))^\top$ ($N=2$) are given by
\begin{equation}\label{pendulum}
\begin{aligned}
\dot{x} &= y, \\
\dot{y} &= -\sin x,
\end{aligned}
\end{equation}
with the Lie derivative (Definition~\ref{def:closed})
\begin{equation}
L = \partial_t + y\,\frac{\partial}{\partial x} - \sin x\,\frac{\partial}{\partial y},
\end{equation}
where $\partial_t$ vanishes on functions of $(x,y)$ alone.

\emph{Rule 3.} As shown in Proposition~\ref{prop:rules}(c) and Remark~\ref{rem:closure}, we introduce $\mathbf{u}=(\sin x, \cos x, y)$ and consider monomials of the form $\sin^a x \cos^b x\, y^{c}$ with $1\le a+b+c=\ell \leq n$, omitting the constant $1$ since no $L\,u_k$ contains a constant term. Placing $g_1=x$ and $g_2=y$ as the first $N=2$ components (fulfilling Definition~\ref{def:closed}(a)) followed by the remaining monomials of $\mathbf{u}$ ordered by degree $\ell$ yields an $L$-closed family $\mathbf{g}$ combining Fourier and Taylor expansions. Notice also that although the solution involves infinite series in all powers of the initial conditions $(x_0, y_0)$, the pendulum is integrable and exhibits regular dynamics.

\begin{figure}[H]
\centering
\scalebox{.8}{\includegraphics{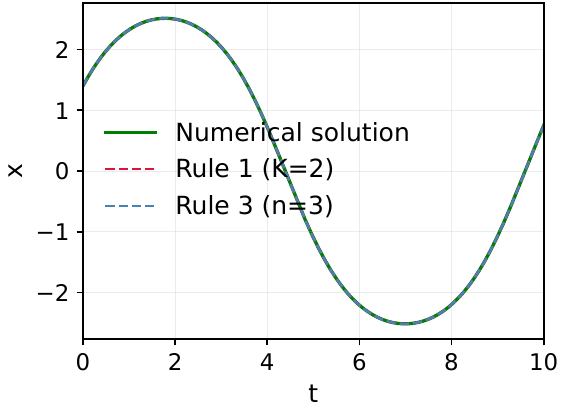}}
\caption{Nonlinear pendulum: time evolution of $x(t)$ obtained with Rule~1 ($K=2$) and Rule~3 (degree $n=3$), together with the numerical solution. The initial conditions are $x(0)=y(0)=1.4$. The pair $(K,n)=(2,3)$ is not matched in Taylor-order information.}
\label{Fig6}
\end{figure}

With initial conditions $x(0)=y(0)=1.4$, we compute the trajectory $\mathbf{x}(t)=(x(t),y(t))^\top$ over the interval $[0,10]$ by concatenating consecutive solutions of equal length $\Delta t=0.01$. Here each application of $L$ raises the degree in $\mathbf{u}$ by at most one ($L\sin x = y\cos x$, $L\cos x = -y\sin x$, $Ly = -\sin x$, and $Lx = y$), so $L^m x$ has degree at most $m-1$ in $\mathbf{u}$ for $m \ge 2$ (specifically, $L^4 x = \sin x\cos x + y^2\sin x$ has degree $3$). Thus, truncating Rule~3 at the lowest degree $n=3$ that contains $L^4x$ ($1 \leq \ell \leq 3$) reproduces exact Taylor coefficients of $(x,y)$ up to order $M(3)=3$ (and up to order $4$ for $x$ alone). This is compared in Fig.~\ref{Fig6} against Rule~1 at temporal order $K=2$ and the reference numerical solution. We stress that, unlike the Lorenz and quartic cases, the pair $(K,n)=(2,3)$ is \emph{not} matched in Taylor-order information: Rule~3 at $n=3$ retains at least one more Taylor order than Rule~1 at $K=2$, so the comparison in Fig.~\ref{Fig6} is purely illustrative.

For illustration at degree $n=2$, the truncated vector of products in $\mathbf{u}=(\sin x,\cos x,y)$ is given by
\begin{equation}\label{pendulo_vector_exten}
\mathbf{u}_{(2)}(t)=\left(\sin x, \cos x, y, \sin^2 x, \sin x\cos x, \sin x\, y, \cos^2 x,\cos x\, y, y^2 \right)^\top,
\end{equation}
which, after placing $(g_1,g_2)=(x,y)$ first (without repeating $y$), forms the truncated extended vector $\mathbf{X}_{(2)}(t)$ (a finite truncation of the $L$-closed family $\mathbf{g}$, which does not close under $L$ by itself) from which $\mathbf{x}(t)=(x(t),y(t))^\top$ on the $\gamma$-th interval is projected via $[e^{\mathbf{A}_{(2)} \tau}]_R\,\bar{\mathbf{X}}_{(2),\gamma}$.

\section{Nonlinearity: a unique route for chaos in ODE systems?}\label{sec:chaos}
In this section, we discuss the role of nonlinearity as a necessary condition for chaos to occur, based on the previous results. First, we note that any real-analytic ODE system can be expressed as an infinite linear ODE system by redefining the variables, as shown in Section~\ref{sec:formalism}. Consequently, when the extended trajectory of an infinite linear system corresponds to a chaotic finite-dimensional nonlinear system, the chaotic dynamics is inherited. We denote by $\Omega$ the finite-dimensional original phase space, built from the variables $\mathbf{x}(t)$ of the nonlinear system~\eqref{sys equat}. In the infinite-dimensional formulation, the extended trajectory $\mathbf{X}(t)=\mathbf{g}(\mathbf{x}(t))$ of consistent initial data stays on the embedded manifold $\{\mathbf{g}(\mathbf{x}) : \mathbf{x} \in \Omega\}$ for as long as the original solution exists, because the algebraic relations among the extended variables are preserved, so that the dynamics restricted to this manifold is conjugate to the original nonlinear system. Since $\mathbf{A}$ is in general unbounded, we formulate this statement on trajectories rather than on a linear flow acting on a Banach space. In this finite-dimensional setting, chaos is characterized in the standard dynamical-systems sense by topological transitivity/mixing on a compact invariant set together with sensitive dependence on initial conditions, quantified through the positivity of Lyapunov exponents on bounded subsets of $\Omega$.

Having established this, one may ask more subtly: could the infinite-dimensional linear system itself exhibit chaotic behavior? Since the solution of an ODE system is determined not only by the equations but also by their initial conditions, we can always construct systems of infinite linear equations based on chaotic nonlinear systems, by considering small changes in the initial conditions within the infinite-dimensional vector $\mathbf{x}_E(t)$ of Eq.~\eqref{X}. Moreover, this can be done in such a way that the system is not reducible to a finite number of equations. It is therefore plausible that these systems behave similarly to the original ones. Consequently, the subvector $\mathbf{x}(t)$ of Eq.~\eqref{X} may exhibit dynamics similar to the original $\mathbf{x}(t)$ and thus appear chaotic. Indicators such as sensitivity to initial conditions or irregular trajectories could also be observed in infinite linear systems, provided the initial conditions are correlated in a particular way.

\begin{remark}\label{rem:sensitivity}
It is worth noting that sensitivity to initial conditions may arise in the infinite linear system even without starting from a chaotic nonlinear system. This can occur because the series of equations may involve correlations among powers of the initial conditions, so that small variations in $\mathbf{x}_0$ may be amplified through higher-order terms. In this sense, the infinite linear formulation may exhibit sensitivity induced by the algebraic structure of the initial data, although whether this qualifies as genuine chaos in the classical sense remains an open question.
\end{remark}

Here we encounter an obvious and debated limitation: what is the phase space? Can we define a phase space of infinite dimensions? May we consider extensions of topological chaos, such as Devaney chaos and hypercyclicity in linear functional spaces~\cite{grosse,godefroy,Bonet}? In principle, the phase space could be defined through infinitely many Poincar\'{e} sections, providing a complete description of the correlations among variables. In practice, one may expect that only a finite number of sections is required to preserve the essential structure of the dynamics; we conjecture that such a finite-section prescription can bridge the gap between the theoretical infinite-dimensional setting and the finite-dimensional approximations used in analysis and computation, but this remains to be established.

Furthermore, what does ``infinite'' mean in practice? Is it different from 1000, 10000, or a macroscopic number of variables? In such cases, one might be interested in only a subset of variables. One could then focus on those variables, regardless of whether the initial conditions of the remaining variables change. More formally, one could also work in a subspace of phase space defined by Poincar\'{e} sections. A detailed exploration of these questions, particularly the operational meaning of ``infinite'' and the role of Poincar\'{e} sections, will be addressed in future work.

\section{Conclusions}\label{sec:conclusions}
In this article, we have introduced a methodology that allows one to find the solution of a real-analytic nonlinear system by rewriting it as an extended system of infinitely many linear equations. This methodology involves considering as new variables the nonlinear functions in the ODE system and their infinite subsequent derivatives. We have shown that the choice of new variables can be made according to different rules or criteria (theoretically infinitely many), determining different $\mathbf{x}_E(t)$ vectors in Eq.~\eqref{X}. In this way, the analytic solution of the system is represented locally through a piecewise-defined function of time and initial conditions~\eqref{pegando}, approximated via finite truncations.

Three rules were applied in three illustrative examples. The qualitative behavior of the trajectories is reproduced in all three examples (Figs.~\ref{Fig1}, \ref{Fig2}, \ref{fig:hamilt_quart}, and~\ref{Fig6}), with quantitative accuracy depending on the system, truncation order, and chosen basis; in particular, for the Lorenz system the error of Rule~2 ($n=5$) is larger than that of Rule~1 ($K=4$), and no error measure is reported for the pendulum. The first rule introduces the successive time derivatives of the state vector as extended variables, namely $\mathbf{X}^{(0)}(t)=\mathbf{x}(t)$, $\mathbf{X}^{(1)}(t)=\mathbf{x}'(t)$, $\ldots$, $\mathbf{X}^{(j)}(t)=\mathbf{x}^{(j)}(t)$ with $j=0,1,2,\ldots$, whose finite truncations at order $K$ yield the $K$-th degree Taylor polynomials in time. This rule was used in all the examples analyzed: the Lorenz model, the quartic classical Hamiltonian, and the nonlinear pendulum. Secondly, in the Lorenz model and the quartic classical Hamiltonian, we considered monomials of the state variables ordered by total degree $\ell=1,2,\ldots$ (e.g., $x^a y^b z^{\ell-a-b}$ in Lorenz, Rule~2, which is Carleman linearization about the equilibrium at the origin), which leads to an expansion in powers of the initial conditions. Finally, for the nonlinear pendulum, we considered mixed products $\sin^a x \cos^b x\, y^{c}$ ($1\le a+b+c=\ell$) together with $x$, combining Fourier and Taylor bases. In all numerical comparisons, a fixed time step $\Delta t = 0.01$ was used so that the trajectories are evaluated on the same temporal mesh, while truncation orders were matched via $n = 1 + K(d-1)$ for polynomial systems of degree $d$ (Rules~1 and~2). For the nonlinear pendulum (Rule~3) the pair $(K,n)=(2,3)$ was used, which is not matched in Taylor-order information, as detailed in Sec.~\ref{sec:pendulum}.

Although in all cases the infinite expansions share the Taylor coefficients of the solution near $t_0$, their finite truncations (partial sums) can yield different accuracy and reveal distinct dynamical structures. For instance, in the Lorenz model, Rule~2 yields real exponentials (Eq.~\eqref{exponential}) instead of powers of $\tau$ as in Rule~1; its lower accuracy for the Lorenz system, where the initial state is far from the equilibrium, may be related to the dependence on the distance to the equilibrium reported for Carleman approximants \cite{MS25}, while in the quartic classical Hamiltonian, for the case studied, Rule~2 ($n=7$) improves both trajectory accuracy and energy conservation compared to Rule~1 ($K=3$). Furthermore, comparing the dependence on the initial conditions across the examples shows that while nonlinear terms generate infinite series in powers of $\mathbf{x}_0$ (unlike finite-dimensional linear systems), the mere presence of such powers is not sufficient for chaos, as illustrated by the regular dynamics of the nonlinear pendulum.

In conclusion, analyzing nonlinear systems from the perspective of infinite linear equations opens several new avenues. One example is the possible chaotic behavior of infinite linear ODE systems, with an appropriate definition of chaos for such functional spaces.

\section*{ACKNOWLEDGMENTS}
The authors would like to thank Dr. Catalina Mostaccio for her collaboration. Ignacio S. Gomez acknowledges support from the Department of Exact and Natural Sciences of the State University of Southwest Bahia (UESB), Itapetinga, Bahia, Brazil, from the PROFÍSICA (UESC), Ilhéus, Bahia, Brazil and from the Conselho Nacional de Desenvolvimento Científico e Tecnológico (CNPq), Grant Number 316131/2023-7.

\appendix
\section{Structure of $\mathbf{A}$}\label{app:struct}
Let the extended variables be grouped by degree $\ell=1,2,\dots$ (the derivative order plus one for Rule~1, and the total degree of the monomials or of the products of $\sin x$, $\cos x$, and $y$ for Rules~2 and~3). We assume that $L$ does not lower the degree. For monomial bases of the state variables this holds when $\mathbf{f}$ has no constant term ($c_{i,\boldsymbol{\kappa}}=0$ for $|\boldsymbol{\kappa}|=0$, see the proof of Proposition~\ref{prop:rules}); otherwise a degree-$0$ variable (the constant $1$) appears and $L\mathbf{x}^{\boldsymbol{\alpha}}$ contains terms of degree $|\boldsymbol{\alpha}|-1$, so that $\mathbf{A}_{\ell,\ell-1}\neq\mathbf{0}$ and the block triangular structure below is lost. Under this assumption, $\mathbf{A}=(\mathbf{A}_{\ell\ell'})$ is \emph{block upper triangular}:
\begin{equation}
\mathbf{A}_{\ell\ell'}=\mathbf{0} \quad \text{for } \ell>\ell' ,
\end{equation}
and each block row has finitely many nonzero blocks. We denote by $\mathbf{A}^{[\ell,\ell']}$ the finite section formed by the block rows and columns $\ell,\dots,\ell'$, and $\mathbf{A}_{(n)}=\mathbf{A}^{[1,n]}$; for Rule~1 at Taylor order $K$ the subindex counts blocks, so that $n=K+1$. Two cases must be distinguished.
\begin{itemize}
\item \emph{Rule 1.} The diagonal blocks vanish, so $\mathbf{A}$ is strictly block upper triangular [Proposition~\ref{prop:rules}(a)].
\item \emph{Rules 2 and 3.} The diagonal blocks $\mathbf{A}_{\ell\ell}$ are finite, nonzero, and in general full; for the Lorenz matrix~\eqref{eq:A2_lorenz}, for instance, $\mathbf{A}_{\ell\ell}$ contains entries below its diagonal ($\rho$ in the first block). Hence $\mathbf{A}$ is block upper triangular, but \emph{not} upper triangular in the strict sense.
\end{itemize}

\section{Well-definedness of the exponential}\label{app:welldef}
The exponential is defined formally by the series
\begin{equation}
e^{\mathbf{A}t}=\sum_{m=0}^{\infty}\frac{(\mathbf{A}t)^m}{m!}.
\label{eq:expdef}
\end{equation}
For an infinite matrix this requires care: the entries of $\mathbf{A}$ grow with the degree (e.g., $-2\sigma$, $-2\beta$ in Eq.~\eqref{eq:A2_lorenz}), so $\mathbf{A}$ is not a bounded operator and the symbolic calculus of a Banach algebra~\cite{Rudin} does not apply directly. The block upper triangular structure makes this unnecessary: the block $(\ell,\ell')$ of any power $\mathbf{A}^m$ is a sum over chains $\ell\le k_1\le\dots\le k_{m-1}\le \ell'$ and involves only the blocks $\mathbf{A}_{kk'}$ with $\ell\le k\le k'\le \ell'$. Therefore,
\begin{equation}
\big(e^{\mathbf{A}t}\big)_{\ell\ell'}=\big(e^{\mathbf{A}^{[\ell,\ell']}t}\big)_{\ell\ell'},
\label{eq:section}
\end{equation}
where $\mathbf{A}^{[\ell,\ell']}$ is the finite section of $\mathbf{A}$ defined in Appendix~\ref{app:struct}. Each block of $e^{\mathbf{A}t}$ is thus the corresponding block of the exponential of a finite matrix, and the series~\eqref{eq:expdef} converges block by block. This defines $e^{\mathbf{A}t}$, and in particular its first $N$ rows, as a formal infinite matrix. The quantity $\big[e^{\mathbf{A}t}\big]_R\,\mathbf{X}(0)$ is defined in Theorem~\ref{thm:main}(iii) by the Taylor series, whose convergence to the solution is guaranteed for $|t|<r(\mathbf{x}_0)$. Its identification with the product of the first $N$ rows of the blockwise exponential and the infinite vector $\mathbf{X}(0)$ requires interchanging the series in $m$ with the infinite sum over the column index (the $i$-th row of $e^{\mathbf{A}t}$ has infinitely many nonzero entries, since the degree grows with $m$); this interchange is justified under absolute convergence of $\sum_m\sum_j |t|^m/m!\,|(\mathbf{A}^m)_{ij}|\,|X_j(0)|$, which we do not prove. Throughout the paper we therefore use the series definition of Theorem~\ref{thm:main}(iii); the behavior of finite truncations is discussed in Lemma~\ref{lem:coef} and Remark~\ref{rem:scope}.

Eq.~\eqref{eq:section} also gives the meaning of the truncation at degree $n$: the finite matrix $\mathbf{A}_{(n)}=\mathbf{A}^{[1,n]}$ yields exactly the contribution of the variables up to degree $n$, and the approximation error is the contribution of the omitted variables of degree larger than $n$.

\section{Nilpotent case (Rule 1)}\label{app:nilp}
In Rule~1 at Taylor order $K$, all diagonal blocks vanish and the truncated matrix $\mathbf{A}_{(K+1)} = \mathbf{A}^{[1,K+1]}$ (of dimension $N(K+1)\times N(K+1)$, retaining the $K+1$ blocks $\mathbf{x}, \mathbf{x}', \dots, \mathbf{x}^{(K)}$) is strictly block upper triangular, hence nilpotent with $\mathbf{A}_{(K+1)}^{\,K+1}=\mathbf{0}$, so that
\begin{equation}
e^{\mathbf{A}_{(K+1)} t}=\sum_{m=0}^{K}\frac{(\mathbf{A}_{(K+1)} t)^m}{m!}.
\label{eq:nilp}
\end{equation}
The infinite matrix itself is not nilpotent (it is a shift), but every entry of $e^{\mathbf{A}t}$ is a polynomial in $t$, which reproduces the Taylor expansion~\eqref{Taylorfinal}. Note that nilpotency depends on the choice of the new variables, not on whether the original ODE system contains linear terms: Rule~1 gives a nilpotent truncation even for the Lorenz system.

\section{General case}\label{app:gen}
When some diagonal block is nonzero, write $\mathbf{A}=\mathbf{A}_{\mathrm{diag}}+\mathbf{A}_{\mathrm{up}}$, with $\mathbf{A}_{\mathrm{diag}}$ the block diagonal part and $\mathbf{A}_{\mathrm{up}}$ the strictly block upper triangular part. If $[\mathbf{A}_{\mathrm{diag}},\mathbf{A}_{\mathrm{up}}]=\mathbf{0}$, then $e^{\mathbf{A}}=e^{\mathbf{A}_{\mathrm{diag}}}e^{\mathbf{A}_{\mathrm{up}}}$. In general $\mathbf{A}_{\mathrm{diag}}$ and $\mathbf{A}_{\mathrm{up}}$ do not commute (this is the case of Rule~2), and the factorization is not valid. Two alternatives are available.

\begin{itemize}
\item \emph{Zassenhaus formula}~\cite{casas}, with $\mathbf{D}\equiv\mathbf{A}_{\mathrm{diag}}$ and $\mathbf{U}\equiv\mathbf{A}_{\mathrm{up}}$ for brevity:
\begin{equation}
e^{\mathbf{D}+\mathbf{U}}=e^{\mathbf{D}}e^{\mathbf{U}}
\exp\!\Big(-\tfrac12[\mathbf{D},\mathbf{U}]\Big)
\exp\!\Big(\tfrac16[\mathbf{D},[\mathbf{D},\mathbf{U}]]
+\tfrac13[\mathbf{U},[\mathbf{D},\mathbf{U}]]\Big)\cdots,
\label{eq:zass}
\end{equation}
where the dots denote exponentials of higher-order nested commutators. (The Baker--Campbell--Hausdorff formula~\cite{hall} gives instead $\log(e^{\mathbf{X}}e^{\mathbf{Y}})$, whose third-order terms have coefficients $1/12$.) Since $[\mathbf{D},\mathbf{U}]\neq\mathbf{0}$ the product of commutator exponentials does not terminate in general, so truncating it is useful mainly as a perturbative expansion.

\item \emph{Direct evaluation.} The spectrum of $\mathbf{A}$ is the union of the spectra of the diagonal blocks. If each $\mathbf{A}_{\ell\ell}$ is diagonalizable, a block diagonal similarity transformation brings $\mathbf{A}$ to upper triangular form $\tilde{\mathbf{A}} = (\tilde{a}_{kl})$ with eigenvalues $\lambda_k$ on the diagonal. The entries of the exponential then follow from confluent divided differences (accounting for repeated eigenvalues, such as $\pm i$ in the quartic Hamiltonian) of the exponential function,
\begin{equation}
\big(e^{\tilde{\mathbf{A}}t}\big)_{kl}=\sum_{m\ge 0}\sum_{k=j_0<j_1<\dots<j_m=l}
\tilde{a}_{j_0j_1}\cdots \tilde{a}_{j_{m-1}j_m}\,
\phi\big[\lambda_{j_0},\dots,\lambda_{j_m}\big],
\qquad \phi(\lambda)=e^{\lambda t},
\end{equation}
where $\phi[\lambda_{j_0},\dots,\lambda_{j_m}]$ denotes the confluent divided difference of $\phi$ at those nodes (the term $m=0$, i.e., $k=l$, is $e^{\lambda_k t}$). This is how exponentials of the eigenvalues, as in Eq.~\eqref{exponential}, appear in the solution for Rule~2.
\end{itemize}

\section{Existence and uniqueness}\label{app:exist}
For every finite truncation $\mathbf{A}_{(n)}$, the linear ODE system has a unique global solution. For the infinite system with unbounded $\mathbf{A}$, the Picard--Lindel\"{o}f theorem and the contraction mapping principle~\cite{Rudin,Burton} do not apply directly, since they require a Lipschitz (bounded) right-hand side. On the manifold of consistent initial data $\mathbf{X}(t_0)=\mathbf{g}(\mathbf{x}_0,t_0)$, a local analytic solution is constructed via the Taylor series~\eqref{Taylorfinal}, whose uniqueness within the class of solutions generated by the flow of~\eqref{sys equat} follows from the uniqueness of solutions to the original nonlinear system~\eqref{sys equat} (whereas uniqueness of arbitrary solutions to the unconstrained infinite-dimensional linear system in sequence spaces is not required by our construction).

\section{Numerical estimation of the Lyapunov exponent}\label{Apendice_Lya}
The maximum finite-time Lyapunov exponent, $\lambda_{\text{L}}(t)$, measures the rate of separation between two trajectories with nearby initial conditions. It is computed as
\begin{equation}
\lambda_{\text{L}}(t) = \frac{1}{t} \ln \left( \frac{\Vert\mathbf{x}_{\text{pert}}(t) - \mathbf{x}_{\text{base}}(t)\Vert_2}{\Vert\mathbf{x}_0^{\text{pert}} - \mathbf{x}_0\Vert_2} \right),
\label{eq:lyapunov}
\end{equation}
where $\Vert\cdot\Vert_2$ denotes the Euclidean norm, $\delta_0 = 10^{-8}$, and $\Vert\mathbf{x}_0^{\text{pert}} - \mathbf{x}_0\Vert_2 = \sqrt{3}\,\delta_0$.

The simulation parameters are:
\begin{itemize}
\item \emph{Base initial condition:} $\mathbf{x}_0 = [-6.0, -10.0, 14.0]^\top$.
\item \emph{Perturbed initial condition:} $\mathbf{x}_0^{\text{pert}} = \mathbf{x}_0 + \delta_0 \mathbf{1}$, with $\mathbf{1} = [1, 1, 1]^\top$.
\item \emph{Time domain:} $t \in [0, 10]$ with constant sampling step $\Delta t = 0.01$ ($N_{\text{steps}} = 1000$ intervals).
\end{itemize}

For the numerical integration of the base and perturbed trajectories, we used as reference the adaptive Runge--Kutta scheme of order 4(5) (the \texttt{RK45} implementation of the \texttt{scipy.integrate.solve\_ivp} function in Python), executed with relative and absolute tolerances $\text{rtol} = \text{atol} = 10^{-12}$ and sampled on the grid $t_k=k\Delta t$. The same perturbation $\delta_0\mathbf{1}$ is applied to the initial condition of the Rule~1 and Rule~2 approximations, which are evaluated with the same step $\Delta t$.

Since at $t = 0$ Eq.~\eqref{eq:lyapunov} presents an indeterminacy of type $0/0$, the value $\lambda_{\text{L}}(0)$ is omitted and the evaluation is performed exclusively for $t > 0$.


\begin{thebibliography}{99}

\bibitem{Coddington}
E. A. Coddington and N. Levinson, \textit{Theory of Ordinary Differential Equations} (McGraw-Hill, New York, 1955).

\bibitem{tabor}
M. Tabor, \textit{Chaos and Integrability in Nonlinear Dynamics: An Introduction} (Wiley, New York, 1988).

\bibitem{lich}
A. J. Lichtenberg and M. A. Lieberman, \textit{Regular and Chaotic Dynamics} (Springer-Verlag, New York, 1992).

\bibitem{Lorenz}
E. N. Lorenz, \textit{J. Atmos. Sci.} \textbf{20}, 130 (1963).

\bibitem{Rossler}
O. E. R\"{o}ssler, \textit{Phys. Lett. A} \textbf{57}, 397 (1976); \textbf{71}, 155 (1979).

\bibitem{Poinc}
H. Poincar\'{e}, \textit{Oeuvres} (Gauthier-Villars, Paris, 1892), Vol. 1; I. Bendixson, \textit{Acta Math.} \textbf{24}, 1 (1901).

\bibitem{Oseledets}
V. I. Oseledets, \textit{Trans. Moscow Math. Soc.} \textbf{19}, 197 (1968); D. Ruelle, \textit{Publ. Math. IH\'{E}S} \textbf{50}, 275 (1979).

\bibitem{QSDT}
I. Gomez and M. Castagnino, \textit{Chaos, Solitons \& Fractals} \textbf{70}, 99 (2015).

\bibitem{LasotaMackey}
A. Lasota and M. C. Mackey, \textit{Chaos, Fractals, and Noise: Stochastic Aspects of Dynamics}, 2nd ed. (Springer, New York, 1994).

\bibitem{Baladi}
V. Baladi, \textit{Positive Transfer Operators and Decay of Correlations} (World Scientific, Singapore, 2000).

\bibitem{Bonet}
J. Bonet, F. Mart\'{\i}nez-Gim\'{e}nez, and A. Peris, \textit{Bull. London Math. Soc.} \textbf{33}, 196 (2001).

\bibitem{Rudin}
W. Rudin, \textit{Functional Analysis} (McGraw-Hill, New York, 1973).

\bibitem{Burton}
T. A. Burton, \textit{Stability by Fixed Point Theorem for Functional Differential Equations} (Dover, New York, 2006).

\bibitem{carleman1932}
T. Carleman, \textit{Acta Math.} \textbf{59}, 63 (1932).

\bibitem{kowalskisteeb1991}
K. Kowalski and W.-H. Steeb, \textit{Nonlinear Dynamical Systems and Carleman Linearization} (World Scientific, Singapore, 1991).

\bibitem{koopman1931}
B. O. Koopman, \textit{Proc. Natl. Acad. Sci. U.S.A.} \textbf{17}, 315 (1931).

\bibitem{williams2015}
M. O. Williams, I. G. Kevrekidis, and C. W. Rowley, \textit{J. Nonlinear Sci.} \textbf{25}, 1307 (2015).

\bibitem{berz}
M. Berz, \textit{Modern Map Methods in Particle Beam Physics} (Academic Press, San Diego, 1999).

\bibitem{galileo}
D. J. Gondelach, R. Armellin, and A. Wittig, arXiv:1901.06947 (2019).

\bibitem{lagrangiandescriptorsDA}
A. C\u{a}liman, J. Daquin, and A. S. Libert, \textit{Physica D} \textbf{472}, 134506 (2025).

\bibitem{taylorchaoticmaps}
H. Richter and G. Stein, \textit{Chaos, Solitons \& Fractals} \textbf{13}, 1783 (2002).

\bibitem{carlemanbounds}
A. Amini, C. Zheng, Q. Sun, and N. Motee, arXiv:2207.07755 (2022).

\bibitem{carlemanfourier}
P. Chen, N. Motee, and Q. Sun, arXiv:2411.11598 (2024).

\bibitem{carlemanglobal}
I. Novikau and I. Joseph, arXiv:2510.15715 (2025).

\bibitem{carlemanradio2016}
H. Weber and W. Mathis, \textit{Adv. Radio Sci.} \textbf{14}, 51 (2016).

\bibitem{koopmanchaosspectrum}
M. Kamb, E. Kaiser, S. L. Brunton, and J. N. Kutz, \textit{SIAM J. Appl. Dyn. Syst.} \textbf{19}, 886 (2020).


\bibitem{MS25} Muñoz-Sánchez, J. C., Elena, S. F., \& Oteo, J. Á. (2025).\textit{Carleman approximants for non-linear differential systems}. Physica Scripta, 100(6), 065231.

\bibitem{MH76} Montroll, E. W., \& Helleman, R. H. (1976, March). On a nonlinear perturbation theory without secular terms ii. Carleman embedding of nonlinear equations in an infinite set of linear ones. In AIP conference Proceedings (Vol. 27, No. 1, pp. 75-110). AIP Publishing.

\bibitem{grosse}
K.-G. Grosse-Erdmann and A. Peris Manguillot, \textit{Linear Chaos} (Springer, London, 2011).

\bibitem{godefroy}
G. Godefroy and J. H. Shapiro, \textit{J. Funct. Anal.} \textbf{98}, 229 (1991).

\bibitem{hall}
B. C. Hall, \textit{Lie Groups, Lie Algebras, and Representations} (Springer, New York, 2015).

\bibitem{casas}
F. Casas, A. Murua, and M. Nadinic, \textit{Comput. Phys. Commun.} \textbf{183}, 2386 (2012).

\end{thebibliography}
\end{document}